
\input phyzzx
\magnification=\magstep1
\def\ie{{\it i.e.}}

\def\oh{{\textstyle {1\over 2}}}

\def\text{\textstyle}

\def\Re{{\rm Re}}
\def\refmark#1{[#1]}                        

\vsize=23.5 true cm
\hsize=15.4 true cm
\predisplaypenalty=0
\abovedisplayskip=3mm plus6pt minus 4pt
\belowdisplayskip=3mm plus6pt minus 4pt
\abovedisplayshortskip=0mm plus6pt
\belowdisplayshortskip=2mm plus6pt minus 4pt
\normalbaselineskip=12pt
\normalbaselines
\noindent

\vsize=23.5 true cm
\hsize=15.4 true cm

\def\tr{\rm Tr}

\Pubnum{PUPT- 1335}
\def\refmark#1{[#1]}                        
\REF\kazmig{ V. Kazakov, and  A. Migdal, {\sl Induced QCD at Large N},
Princeton preprint
PUPT-1322, May 1992.}
\REF\mig{ A. Migdal, {\sl Exact Solution of Induced Lattic Gauge Theory
at Large N}, Princeton preprint PUPT-1323, June 1992.}
\REF\migtwo{ A. Migdal, {\sl ${1 \over N}$ Expansion and Particle Spectrum
in Induced QCD,}
Princeton preprint PUPT-1332, June 1992.}
\REF\kogan{ I. Kogan, G. Semenoff and N. Weiss,
{\sl Induced QCD and Hidden Local $Z_N$ Symmetry},
UBCTP 92-022, June 1992.}
\REF\itz{ C. Itzykson and J. Zuber, {\it Jour. Math. Phys. {\bf 21}}, 411
(1980).}
\REF\grosskleb{ D. Gross and I. Klebanov, {\sl One Dimensional String
Theory on a Circle},
{\it Nucl. Phys. \bf B334}, 475 (1990).}

\REF\brez{ E. Brezin, C. Itzykson, G. Parisi and J. Zuber,
{\it Comm. Math. Phys.}  {\bf 59}, (1978) 35.}

\REF\private{ A. Migdal, private communication.}

\titlepage
\title{  {\bf SOME REMARKS ABOUT INDUCED QCD}}
\author{DAVID J. GROSS\
\foot{Research supported in part by NSF grant
PHY90-19754} } \address{\JHL}

\abstract
Migdal and Kazakov have suggested that lattice QCD
with an adjoint representation scalar in the infinite coupling limit could
induce QCD.
I find an exact saddlepoint of this theory for infinite $N$ in the case of a
quadratic scalar potential. I discuss some aspects of this solution
and also  show how the continuum $D=1$ matrix model with an arbitrary potential
can be reproduced through this approach.

\vsize=23.5 true cm
\hsize=15.4 true cm
\chapter{ \bf   Introduction}

Recently Kazakov and Migdal have proposed a  lattice gauge model that,
on the one hand, is  potentially soluble for large N and on the other hand
might
induce QCD \refmark{\kazmig}.
Their idea was to consider a scalar field, $\Phi$,  in the
adjoint representation  of $SU(N)$, on  a lattice of spacing a, coupled to a
$SU(N)$ lattice gauge field in the standard fashion,
$$ S= \sum_{x}  N  \tr [ -U(\Phi(x)) + \sum_{\mu }  \Phi(x)
U_{\mu}(x)  \Phi(x+\mu a) U^{\dagger}_{\mu}(x) ]\,\,,
\eqn\aa$$
($\mu$ runs over the $2D$ lattice vectors on a hypercubic lattice of dimension
$D$.)
Note that there is no kinetic term for the gauge field. This is
equivalent to the infinitely strong coupling limit of the standard lattice
action. They argued that if one integrates out the scalar mesons (even in the
case
of noninteracting scalars with $U(\Phi) = \oh m^2 \Phi^2$),
then at distances large compared to $a$, one would induce in four
dimensions an effective gauge interaction, ${{N\over 48 \pi^2}
(\ln{1\over ma}}) \tr F_{\mu \nu} F^{\mu \nu}$.
The basic idea is that the infrared slavery of the scalars, at the
size of the lattice spacing, produces an effective
gauge theory at a larger scale (much larger
than the inverse scalar mass), which then produces
the usual asymptotically free fixed point
theory. Migdal has analyzed  the large N saddlepoint equations of this model in
great
detail \refmark{\mig} and discussed the
scaling properties of purported solutions,  as well as the
equations for the spectrum of states that emerges \refmark{\migtwo}.

There are many problems with this idea. For one the hard gluons are
not absent and their contribution will overwhelm  that of the
scalars at short distances. Their
asymptotic freedom is  more powerful than the infrared slavery of
scalars. It might be possible to overcome this problem by considering a general
potential
$U(\Phi)$ and fiddling with its parameters to  arrive at the QCD fixed point.
Another issue is that the above theory possesses a much larger symmetry than
the
$SU(N)$ gauge symmetry of the usual lattice action. It is not difficult to see
that, in D dimensions, it is invariant under $(D-1)\times (N-1)$ extra local
$U(1)$-gauge    symmetries.  This is because the transformation
$U_{\mu}(x)\!\! \to   V^{\dagger}_{\mu}(x)U_{\mu}(x)V_{\mu}(x+\mu a)$,
leaves the action invariant as long as $V_{\mu}(x)$ is   a unitary matrix
that commutes with $\Phi(x)$.  If  $V_{\mu}(x)$ were independent of
$\mu$ then this would be the ordinary gauge invariance. Thus we have $D-1$
new gauge symmetries, which are of course isomorphic to the special unitary
transformations that commute with $\Phi$. Thus  $V_{\mu}(x)=
D_{\mu}(x)\Omega (x)$, where $\Omega (x)$ is the unitary matrix that
diagonalizes $\Phi$ and $D_{\mu}(x)$ is diagonal.

This extra symmetry is evident if one considers the naive continuum limit of
\aa,
by expanding about $U_\mu = 1 + i a A_\mu$, since the action will not depend on
the
diagonal components of the vector potential (in a basis where $\Phi$ is
diagonal).
In fact, one must fix, in addition to the usual gauge fixing, the
extra $(D-1)\times (N-1)$ local symmetries to eliminate these components.
This can easily be done.

A subset of this symmetry is the local $Z_N$ symmetry, $U_{\mu}(x) \!\! \to
Z_\mu
U_{\mu}(x)
Z_\mu^{\dagger}$, where $Z_\mu$ is an element of the center  of the group.
This has been noted recently by
Kogan et.al. \refmark{\kogan}.
This symmetry alone prevents the Wilson loop from acquiring an expectation
value, and
should be broken if we are to recover the QCD fixed point from this
formulation.
Presumably the same is true of the extra continuous symmetries discussed above.

In this paper I shall construct an exact solution of the saddlepoint equations
for the
case of a quadratic potential. Remarkably, it turns out that the eigenvalue
distribution, for large
N, is given by a semi-circular law with a mass parameter that
depends on $m$ and on
the dimension of space-time $D$. This trivial solution bears no resemblence to
continuum QCD. Instead it describes the strong coupling lattice theory--
with color singlet mesons.

\chapter{\bf The Quadratic Potential}

The model described by \aa~ can be reduced to an integral over
N degrees of freedom per site,
by integrating out the matrices that diagonalize  $\Phi(x)$.
This can be done, link by link, in
terms of the Itzykson-Zuber integral ($\Delta(\Phi)= \prod_{i<
j}[\Phi_i-\Phi_j]$)
\refmark{\itz},
$$ I(\Phi, \Psi) \equiv \int {\cal D} U e^{N \tr[ \Phi U \Psi U^{\dagger}]} =
c {\det \{ e^{N \Phi_i \Psi_j}\} \over \Delta(\Phi)\Delta(\Psi)} \,\,. \eqn\ba
$$
The effective action for the eigenvalues is then,
$$ S=\!\!\ \sum_{x,i}  [-N U(\Phi_i(x))]  +\!\!
\sum_{x,i\neq j} \ln |\Phi_i(x)-\Phi_j(x)| + \!\!
\sum_{x, \mu} \ln[I(\Phi(x), \Phi(x+\mu a)] .  \eqn\bb
$$
In the large N limit the integral will be dominated by a
translationally invariant saddlepoint
for the density of eigenvalues, $\rho(\nu)$, of the matrices $\Phi$.
The saddlepoint equation is
$$ P\int d \nu {\rho(\nu) \over \Phi_a - \nu} = \oh U'(\Phi_a ) -
D{\rm Lim}_{N\to \infty} \bigl[{\partial
I(\Phi , \Psi ) \over \partial \Phi_a}\bigr]|_{\Phi = \Psi} \,\,. \eqn\op
$$
Migdal has simplified these using the Schwinger-Dyson equations that are
satisfied by $I(\Phi,
\Psi)$ \refmark{\mig}. These are consequences of the fact that $I$ satisfies
$\tr[({1\over N}{\partial \over \partial\Phi})^k] I = \tr (\Psi)^k I$.
The net result is that one
derives an   equation  for the function  $V'(z)\equiv  \int dz {\rho(\nu) \over
z-\nu}$,
whose imaginary part is ${\rm Im} V'(\nu) = - \pi \rho(\nu)$,
$$
{\rm Re} V'(\lambda)  =
P \int {d\nu \over 2 \pi i} \ln [{ { \lambda -
-{1 \over 2D} U'(\nu) - {D-1 \over D} \Re V'(\nu)+i\pi \rho (\nu)} \over   {
\lambda -
-{1 \over 2D} U'(\nu)- {D-1 \over D} \Re V'(\nu) -i\pi \rho (\nu)} }  ].
\eqn\gg
 $$

Migdal has studied the scaling properties of conjectured solutions of this
equation.
Here we
shall find an exact solution in the case of a quadratic,
Gaussian, potential, $U(\Phi) = \oh m^2
\Phi^2$. First, we note that this equation simplifies dramatically for $D=1$.
This is not
surprising since in one dimension the gauge field plays no role, can be
reabsorbed into a
definition of $\Phi$, and \aa~reduces to the standard action for a scalar
field in one dimension. In
particular for the quadratic potential the path integral is a
Gaussian,
$$Z= \int \prod_n {\cal D} \Phi_ne^{ -  N \sum_n  {\rm Tr} \{  {m^2\over 2}
\Phi_n^2-
\Phi_n \Phi_{n+1} \}}\,\,.  \eqn\mm
$$
Thus the eigenvalues of $\Phi$ will be given by the semi-circular distribution,
namely $\pi  \rho(\nu) =  \sqrt{\mu - {\mu^2 \nu^2 \over 4}}$,
where $\mu$ is determined by
the mean of the squares of the eigenvalues,
$\langle {1\over N} {\rm Tr}(  \Phi^2  )\rangle =
{1\over \mu}$.  It is therefore sufficient to calculate the expectation
value of  $ {1\over N} {\rm Tr}(   \Phi^2  ) $,
which is given by the one loop integral,
$$ {1\over N} {\rm Tr}(   \Phi^2  )=
\int_{-\pi}^{\pi}  {d p\over 2 \pi} {1 \over   m^2 +2{\rm cos }p  }
= {1\over \sqrt{m^4-4}}\equiv {1 \over\mu}\,\,.  \eqn\nn $$
It is easy to verify that this solves \gg, using the fact that
$$ V'(z) = { \mu z \over 2} - \sqrt{  {\mu^2 z^2 \over 4}-\mu};
\,\,\, \Re V'(\nu) = \oh \mu \nu , \,
\, {\rm for} \,\, |\nu| \leq {2\over \sqrt{\mu}}  \,\, . \eqn\jk
$$
However, if we return to \gg, we see that the integral
involved is of the same form for any $D$,
as long as $\Re V'(\nu)$ is linear in $\nu$.
This suggests that we can find a solution of
\gg~with a semi-circular distribution of eigenvalues for a quadratic potential
in any
dimension.

We use the result that,
$$
I= \int_{-2\over \sqrt\mu }^{2\over \sqrt\mu } {d\nu \over 2 \pi
i} \ln [{ { \lambda -  {b\over 2} \nu +i\sqrt{\mu - {\mu^2\over 4} \nu^2}}
\over
{\lambda -  {b\over 2} \nu - i \sqrt{\mu - {\mu^2\over 4} \nu^2} }} ]
=\!\! {2\mu  \over (b^2 -\mu^2)}\bigl[\lambda - \sqrt{\lambda^2
- { (b^2 -\mu^2)\over \mu }}\bigr] .  \eqn\jj
$$
This result is easily established
(change variables $\nu = -i{(z-{1\over z}) \over \sqrt{\mu}}$, integrate once
by parts
and then the $z$ integral can be done by contour integration).
In our case, assuming the  semi-circular distribution in \jk,  $ b= m^2 +
{(D-1) \over D}\mu $. On the other hand, from the integral it must be that
$b^2 = \mu^2 + 4 $.
Therefore we derive a quadratic equation for $\mu$, whose solutions are,
$$
\mu_{\pm}(D) = {m^2(D-1) \pm D\sqrt{ m^2 -4(2D-1)} \over 2D-1 }\,\,\, . \eqn\pp
$$
In particular for $D=1$, $\mu_+(1) = \mu = \sqrt{m^4-1}$, which agrees with
the direct calculation of \nn.

Thus we have found a saddlepoint for any $D$, as long as the potential is
quadratic.
It is given by \jk, with $\mu$ given above.
If we start with a large mass, $m^2 > m_c^2=2D$, then only $\mu_+$ is positive
and must be chosen. This solution can be verified to satisfy the large mass
expansion
discused by Kazakov and Migdal \refmark{\kazmig}. I have explicitly checked
that a
semi-circular distribution of eigenvalues, with $\mu =\mu_+=
m^2 - {2D\over m^2} - {2D(2D-1)\over m^6} -{4D(2D-1)^2 \over m^{10}} -\dots$,
satisfies the ${1\over m}$ expansion up to terms of order $({1\over m})^{14}$.

The free energy, $F= -{1\over N^2 {\rm Vol.}} \ln Z$, can be
calculated  at the saddlepoint
by the following device. We use the fact that for $D=1$  the free energy
can be evaluated by Gaussian integration to yield $F(D=1) =
\oh \int_{- \pi}^{\pi} {dp\over 2 \pi} \ln[m^2+ 2 \cos p]=
\oh \ln ({m^2 + \sqrt{m^4 -4}\over 2})$. Then we note that for any $D$
the free energy is given at the large N saddelpoint by,
$$
F(D) = {m^2\over 2N} \tr \Phi_s^2 - {1\over N^2}
\Delta^2(\Phi_s)- {D\over N^2} \ln I(\Phi_s,\Phi_s)
\,\, , \eqn\mj
$$
where $\Phi_s$ is the saddlepoint scalar field. The first two terms are easily
evaluated
for a semi-circular distribution of eigenvalues, which allows us to express
the logarithm of the Itzykson-Zuber integral, for a semi-circular
distribution of eigenvalues with parameter $\mu$, as \nextline
${1\over N^2}\ln I(\mu)= { \sqrt{\mu^2 +4} - \mu \over 2 \mu}
- \oh \ln \bigl({ \sqrt{\mu^2 +4} + \mu \over 2 \mu} \bigr).$
It then follows that the free energy, for a semi-circular
distribution of eigenvalues with paramter $\mu$ is
$$
F(D) = \oh {m^2\over 2 \mu}  + \oh \ln \mu - {D\over 2} \bigl[
\sqrt{1 + {4 \over \mu^2}} -1 - \ln \bigl( \oh + \oh \sqrt{1 + {4 \over
\mu^2}}\bigr)
\bigr] \,\, . \eqn\mk
$$
It is easily checked that the saddlepoint of \mk~determines $\mu$ to satisfy
\pp.

For large mass it is easily verifed that $F$ is bounded from below, and
achieves its minimum at $\mu=\mu_+$.
For $D\leq 1$, $\mu_+$ vanishes at $m^2 = 2D$, and thus we can take a
continuum limit of the theory. For $D>1$, $\mu_+$ is always non-vanishing.
Instead as $m^2 \to 2D$ the other branch $\mu_-$ vanishes. However, we must
approach
the critical value of $m^2$ from below, since only then is $\mu_-$ positive.
If we examine the
free energy in this case we find that it is unbounded from below. Indeed
$ F(D) \buildrel {\mu \to 0} \over \rightarrow  {m^2 - 2D\over \mu}$.
The saddlepoint at $\mu= \mu_-$, which vanishes as $m^2 \to 2D-0$,
is a local {\em maximum}. Thus although we may construct a
continuum limit about
this point it might contain tachyonic excitations.

It is not surprising that $F$ is unbounded from below, for $m^2 < 2D$, since
the action can be bounded, for translationally invariant $\Phi$, by
$ S = N {\rm Vol.}\, \tr [-{m^2\over 2} \Phi^2 + \sum_\mu \Phi U_\mu \Phi
U^{\dagger}_\mu ] > N {\rm Vol.}\, \tr [-{m^2\over 2} \Phi^2 +D  \Phi  \Phi]$,
which is unbounded from below. Thus, this is not a stable region of the
parameters.
Unless  there exist other, healthier, large $N$ saddlepoints of the action,
then there
there would not appear to be a scaling solution that
would give a continuum theory
for  a Gaussian potential for $D>1$. Nonetheless, the saddlepoint we have found
will describe the  large N, strong coupling limit of QCD with adjoint matter.

\chapter{\bf Arbitrary Potential in One Dimension }

Now let us consider the one dimensional system for an arbitrary potential,
namely the model described by

$$Z= \int \prod_n {\cal D} \Phi_ne^{ - \sum_n  {\rm Tr} \{  {m^2\over 2}
\Phi_n^2
+ U(\Phi_n) -
\Phi_n \Phi_{n+1} \}} \,\,. \eqn\oo
$$

This model is hard to solve, except in the double scaling limit, where it was
solved  exactly as long as the spacing between
the points is not too big \refmark{\grosskleb}.
However, in the continuum limit, the model  can be solved in terms of $N$
free fermions in the potential
$ \oh m^2 \Phi^2 + V(\Phi)$, and the density of eigenvalues is
easily calculable in the large $N$ limit.
To take the continuum limit we must scale,
$$  \Phi_n \to {\varphi(t) \over \sqrt{a}}; \,\, \,\, m^2 = 2 + a^2 M^2;
\,\, \,\, U(\Phi)= a W( \varphi) \,\,. \eqn\qq
$$
Then the above partition function reduces, as the lattice spacing $a \to 0$,
to the continuum integral,
$$Z= \int {\cal D} \varphi(t) e^{ - N\int dt\bigl[    {\rm Tr} \{  {M^2\over 2}
\varphi(t)^2
+ W(\varphi(t)) +\oh (\partial_t \varphi(t))^2 \bigr] }  \,\, . \eqn\rr
$$
Now, this model is easily solved, following \refmark{\brez},
by reducing it to the problem of $N$ free fermions, (the eigenvalues of
$\varphi$),
with the one body Hamiltonian, $H= {p^2 \over 2} + {M^2 \over 2}\lambda^2
+W(\lambda)$,
with ${1\over N}$ playing the role of Planck's constant. In the
WKB approximation (\ie~$N\to \infty$ limit),
$$\eqalign{ \rho(E)& ={1\over N} {d n\over  dE}=\int {dx dp \over 2\pi }
\delta [E-{p^2 \over 2} - {M^2 \over 2}x^2 -W(x)] \cr & \rightarrow
\rho(x) = \sqrt{2(E-  {M^2 \over 2}x^2 -W(x))} \,\, . \cr }\eqn\ss
 $$
The Fermi level, $E$, is determined by the condition that $\int_{-x_t}^{x_t}dx
\rho(x)
=1$.

 Let us return to the matrix chain, and
write the equations for this case,

$$\eqalign{
{\rm Re} V'(\lambda)& =
\int {d\nu \over 2 \pi i} \ln [{ { \lambda -(1+ {a^2 M^2 \over 2})\nu
-\oh  {a}^{3\over 2} W'(\sqrt{a}\nu) +i\pi \rho (\nu)} \over{ \lambda -(1+ {a^2
M^2 \over 2})\nu
-\oh  {a}^{3\over 2} W'(\sqrt{a}\nu) -i \pi
\rho (\nu)} } ]\cr {\rm Im }V'(\nu)& = - \pi \rho(\nu) \cr}\,\, .  \eqn\gg
$$

Now rescale the variables so that $\lambda = {\tilde \lambda \over \sqrt{a}},
\,\, \nu= {\tilde \nu\over \sqrt{a}}, \,\, \rho(\nu)\equiv \sqrt{ a f(\tilde
\nu)},$
\noindent $V'(\lambda)\equiv a F(\tilde \lambda)$.
In terms of these the equations read,

$$\eqalign{
{\rm Re} F(\tilde \lambda)& = {1 \over a}
\int {d\tilde\nu \over 2 \pi i} \ln \bigl[{
{\tilde \lambda -\tilde\nu -
{a^2 M^2 \over 2}\tilde\nu
-\oh  {a}^{2} W'(\tilde\nu) +ia \pi f(\tilde\nu)}
\over
{\tilde \lambda -\tilde\nu -
{a^2 M^2 \over 2}\tilde\nu
-\oh  {a}^{2} W'(\tilde\nu) -ia \pi f(\tilde\nu)}
} \bigr]\cr {\rm Im} F(\tilde \nu)& = - \pi f(\tilde \nu) \cr}\,\,. \eqn\uu
$$

Expanding \uu~ in powers of $a$ we find
$$\eqalign{
{\rm Re} F(\tilde \lambda)& =
\int {d\tilde\nu}  { f(\tilde \nu)\over \tilde \lambda - \tilde \nu}
+ {a^2 \over 2}\int d\tilde\nu  {[M^2 \tilde \nu + W'(\tilde \nu)] f(\tilde
\nu)
\over (\tilde \lambda-\tilde \nu)^2}  \cr -&{\pi^2 a^2 \over 3}
\int {d\tilde\nu} {f^3(\tilde \nu) \over (\tilde \lambda-\tilde \nu)^3}
 +O(a^3)
\,\, .\cr } \eqn\ww
$$
Now as $a \to 0$ we see that these equations are trivially satisfied
for {\em any} function $f(\tilde \nu)$.
We now demand that the term of order
$a^2$ vanish. Integrating by parts the last term
this condition can be written as

\noindent
${a^2 \over 2}
\int {d\tilde\nu \over(\tilde \lambda-\tilde \nu)^2 }\bigl(
[M^2 \tilde \nu + W'(\tilde \nu)] f(\tilde \nu) + \pi^2 f^2(\tilde
\nu)f'(\tilde \nu)\bigr)$.
Consequently $
M^2 \tilde \nu + W'(\tilde \nu) + \pi^2 f(\tilde \nu)f'(\tilde \nu)=0$,
an equation that we can solve for $f$,
$$
\pi f(\tilde \nu)= \sqrt{ 2E-M^2{\tilde \nu}^2 -2 W(\tilde \nu)} \,\,. \eqn\xx
$$
$E$ is a constant that can be fixed from the normalization of $\rho$.
This agrees with the fermionic solution \ss.

\chapter{\bf Conclusions}

We have found that infinite coupling QCD, with scalar adjoint matter, is
dominated for large N by a very simple saddlepoint. The eigenvalues of the
scalar field
are distributed according to the semi-circular law. This is quite remarkable,
especially
as the effective scalar action is  certainly not a Gaussian. It will be very
interesting to
evaluate the fluctuations about his saddlepoint--\ie~to work out the effective
Lagrangian describing the confined mesons in this lattice gauge theory.
We note that for zero gauge coupling the eigenvalues also have a
semi-circular distribution. In this case $\mu$ will be given by the analog,
in dimension $D$, of \nn. The fact that the eigenvalue distribution is
semi-circular
for both zero and infinite coupling makes one wonder whether this could be the
case
for finite coupling as well!

What have we learned about induced QCD? We have certainly learned that the
program cannot work for  a Gaussian potential. This is  consistent with the
analysis
of the scaling solutions of the genral equations \refmark{\private}.
So if the idea is to work one must consider non-trivial potentials and
hope that the non-asymptotically free scalar interactions do indeed induce QCD.
Although I do not see  why this could not happen  the physical mechanism
is mysterious.

\refout

\end